\documentclass[%
 reprint,
superscriptaddress,
 amsmath,amssymb,
 aps,
 prl,
]{revtex4-2}

\usepackage{graphicx}
\usepackage{float}
\usepackage{graphicx}
\usepackage{dcolumn}
\usepackage{bm}
\usepackage{verbatim}
\usepackage[dvipsnames]{xcolor}
\usepackage{hyperref}
\usepackage[mathlines]{lineno}
\usepackage{makecell}
\usepackage{ulem}
\usepackage{comment}
\begin{document}

\preprint{APS/123-QED}
\title{Quantum Light Generation based on GaN Microring towards Fully On-chip Source}
\author{Hong Zeng}
\thanks{Hong Zeng, Zhao-Qin He, and Yun-Ru Fan contributed equally to this work.}
\affiliation{Institute of Fundamental and Frontier Sciences, University of Electronic Science and Technology of China, Chengdu 611731, China}
\affiliation{Key Laboratory of Quantum Physics and Photonic Quantum Information, Ministry of Education, University of Electronic Science and Technology of China, Chengdu 611731, China}
\author{Zhao-Qin He}
\thanks{Hong Zeng, Zhao-Qin He, and Yun-Ru Fan contributed equally to this work.}
\affiliation{Department of Electronic Engineering, Tsinghua University, Beijing 100084, China}
\author{Yun-Ru Fan}
\thanks{Hong Zeng, Zhao-Qin He, and Yun-Ru Fan contributed equally to this work.}
\email{Corresponding author: yunrufan@uestc.edu.cn}
\affiliation{Institute of Fundamental and Frontier Sciences, University of Electronic Science and Technology of China, Chengdu 611731, China}
\affiliation{Key Laboratory of Quantum Physics and Photonic Quantum Information, Ministry of Education, University of Electronic Science and Technology of China, Chengdu 611731, China}
\author{Yue Luo}
\affiliation{Institute of Fundamental and Frontier Sciences, University of Electronic Science and Technology of China, Chengdu 611731, China}
\author{Chen Lyu}
\affiliation{Institute of Fundamental and Frontier Sciences, University of Electronic Science and Technology of China, Chengdu 611731, China}
\author{Jin-Peng Wu}
\affiliation{Institute of Fundamental and Frontier Sciences, University of Electronic Science and Technology of China, Chengdu 611731, China}
\affiliation{Key Laboratory of Quantum Physics and Photonic Quantum Information, Ministry of Education, University of Electronic Science and Technology of China, Chengdu 611731, China}
\author{Yun-Bo Li}
\affiliation{\mbox{Department of Fundamental Network Technology, China Mobile Research Institute, Beijing 100053, China.}}
\author{Sheng Liu}
\affiliation{\mbox{Department of Fundamental Network Technology, China Mobile Research Institute, Beijing 100053, China.}}
\author{Dong Wang}
\affiliation{\mbox{Department of Fundamental Network Technology, China Mobile Research Institute, Beijing 100053, China.}}
\author{De-Chao Zhang}
\affiliation{\mbox{Department of Fundamental Network Technology, China Mobile Research Institute, Beijing 100053, China.}}
\author{Juan-Juan Zeng}
\affiliation{Institute of Fundamental and Frontier Sciences, University of Electronic Science and Technology of China, Chengdu 611731, China}
\affiliation{Center for Quantum Internet, Tianfu Jiangxi Laboratory, Chengdu 641419, China}
\author{Guang-Wei Deng}
\affiliation{Institute of Fundamental and Frontier Sciences, University of Electronic Science and Technology of China, Chengdu 611731, China}
\affiliation{Key Laboratory of Quantum Physics and Photonic Quantum Information, Ministry of Education, University of Electronic Science and Technology of China, Chengdu 611731, China}
\author{You Wang}
\affiliation{Institute of Fundamental and Frontier Sciences, University of Electronic Science and Technology of China, Chengdu 611731, China}
\affiliation{Southwest Institute of Technical Physics, Chengdu 610041, China}
\author{Hai-Zhi Song}
\affiliation{Institute of Fundamental and Frontier Sciences, University of Electronic Science and Technology of China, Chengdu 611731, China}
\affiliation{Southwest Institute of Technical Physics, Chengdu 610041, China}
\author{Zhen Wang}
\affiliation{\mbox{National Key Laboratory of Materials for Integrated Circuits, Shanghai Institute of Microsystem and Information Technology,} \mbox{Chinese Academy of Sciences, Shanghai 200050, China}}
\author{Li-Xing You}
\affiliation{\mbox{National Key Laboratory of Materials for Integrated Circuits, Shanghai Institute of Microsystem and Information Technology,} \mbox{Chinese Academy of Sciences, Shanghai 200050, China}}
\author{Kai Guo}
\email{Corresponding author: guokai07203@hotmail.com}
\affiliation{Institute of Systems Engineering, AMS Beijing 100141, China}
\author{Chang-Zheng Sun}
\email{Corresponding author: czsun@tsinghua.edu.cn}
\affiliation{Department of Electronic Engineering, Tsinghua University, Beijing 100084, China}
\author{Yi Luo}
\affiliation{Department of Electronic Engineering, Tsinghua University, Beijing 100084, China}
\author{Guang-Can Guo}
\affiliation{Institute of Fundamental and Frontier Sciences, University of Electronic Science and Technology of China, Chengdu 611731, China}
\affiliation{Key Laboratory of Quantum Physics and Photonic Quantum Information, Ministry of Education, University of Electronic Science and Technology of China, Chengdu 611731, China}
\affiliation{Center for Quantum Internet, Tianfu Jiangxi Laboratory, Chengdu 641419, China}
\affiliation{CAS Key Laboratory of Quantum Information, University of Science and Technology of China, Hefei 230026, China}
\author{Qiang Zhou}
\email{Corresponding author: zhouqiang@uestc.edu.cn}
\affiliation{Institute of Fundamental and Frontier Sciences, University of Electronic Science and Technology of China, Chengdu 611731, China}
\affiliation{Key Laboratory of Quantum Physics and Photonic Quantum Information, Ministry of Education, University of Electronic Science and Technology of China, Chengdu 611731, China}
\affiliation{Center for Quantum Internet, Tianfu Jiangxi Laboratory, Chengdu 641419, China}
\affiliation{CAS Key Laboratory of Quantum Information, University of Science and Technology of China, Hefei 230026, China}

\date{\today}
\begin{abstract}
Integrated quantum light source is increasingly desirable in large-scale quantum information processing.~Despite recent remarkable advances, new material platform is constantly being explored for the fully on-chip integration of quantum light generation, active and passive manipulation, and detection. Here, for the first time, we demonstrate a gallium nitride (GaN) microring based quantum light generation in the telecom C-band, which has potential towards the monolithic integration of quantum light source.~In our demonstration, the GaN microring has a free spectral range of 330 GHz and a near-zero anomalous dispersion region of over 100 nm. The generation of energy-time entangled photon pair is demonstrated with a typical raw two-photon interference visibility of 95.5$\pm$6.5\%, which is further configured to generate heralded single photon with a typical heralded second-order auto-correlation $g^{(2)}_{H}(0)$ of 0.045$\pm$0.001. Our results pave the way for developing chip-scale quantum photonic circuit.                                                                                                                                                                                                                             

\end{abstract} 
\maketitle
\textit{Introduction.}—Quantum photonic integrated circuit (QPIC) provides a promising approach to developing future nonclassical technologies\cite{lu2021advances, pelucchi2022potential, wang2020integrated},~which is considered one of the most competitive candidates for the scalable implementation of quantum communication,~quantum metrology,~quantum simulation and computation\cite{simon2017towards, xu2020secure, hu2023progress, fitzke2022scalable}.~Advances in complementary metal-oxide semiconductor (CMOS) fabrication enable the functionality of tabletop quantum optics to be scaled down to prototype chips with significant improvements in efficiency, robustness, and stability\cite{moody20222022}.~For instance, quantum light generation\cite{sciara2021scalable, wang2018multidimensional}, quantum photonic storage\cite{liu2020demand, wallucks2020quantum, xueyingzhang2023}, and single-photon detection\cite{arrazola2021quantum} have been realized in chip-to-chip quantum teleportation\cite{llewellyn2020chip}, quantum key distribution\cite{sibson2017chip, zhang2019integrated,semenenko2020chip}, and quantum boson sampling\cite{spring2013boson, tillmann2013experimental, paesani2019generation, madsen2022quantum}.~Recently, the integration of quantum light generation with active and passive manipulations has been demonstrated\cite{elshaari2020hybrid, mahmudlu2023fully} with low-loss indirect-bandgap materials and direct-bandgap III-V semiconductors, such as silica\cite{reimer2016generation,kues2017chip}, silicon\cite{ma2017silicon,guo2018generation, liu2020entanglement,liu202240}, silicon nitride\cite{imany201850,wu2021integrated,wen2022realizing,fan2023multi,wen2023polarization}, lithium niobate\cite{ma2020ultrabright, zhao2020high}, gallium aluminum arsenide\cite{steiner2021ultrabright}, indium phosphide\cite{kumar2019entangled}, aluminum nitride\cite{guo2017parametric}, and silicon carbide\cite{ma2023entangled}.~The indirect-bandgap material with high refractive index is usually used for light guiding and entangled photon pair generating, while the direct-bandgap III-V semiconductor is suitable for optical gain and lasing.~Despite these advances, it remains challenging to combine different blocks to build a complex quantum circuit on single chip, which is primarily due to the absence of a favorable quantum material platform encompassing all required functionalities\cite{elshaari2020hybrid, mahmudlu2023fully}.

As a III-V semiconductor with a bandgap of 3.4 eV, gallium nitride (GaN) is a promising material for next-generation photonic and electronic devices.~It exhibits a wide optical transparency window extending from ultra-violet to mid-infrared wavelength.~Its non-centrosymmetric crystal structure endows both the second-order and the third-order nonlinearities\cite{stassen2019high, munk2018four}. These features, coupled with the epitaxial growth of GaN on sapphire (GaNOI) and the recent demonstration of a low-loss GaNOI integrated photonics platform, exemplified by the generation of the second harmonics and the Kerr combs\cite{Xiong:11, moody2020chip, zheng2022integrated}, highlight remarkable capabilities.~Furthermore, the GaN has emerged as a unique quantum material for single-photon emitter known as quantum dot or defect center at room temperature\cite{arita2017ultraclean, stachurski2022single, kako2014single, yuan2023gan, luo2020quantum, kako2006gallium, berhane2017bright, sun2019single, meunier2023telecom}.~Leveraging its favorable characteristics for optical gain and lasing, the GaN also exhibits an excellent potential towards fully integrated quantum photonic circuit.~These advantages inspire us to demonstrate the generation of quantum light based on the spontaneous four-wave mixing (SFWM) process in GaN, the verification of which is urgently expected to pave the way for the development of the QPIC.

In this Letter, we demonstrate the generation of correlated/entangled photon pairs in a GaN microring resonator (MRR) via the SFWM process for the first time. In our experiments, the GaN MRR is designed with a free spectral range (FSR) of 330 GHz and near-zero anomalous dispersion over 100 nm in the telecom C-band. In our demonstration, correlated photon pairs are generated within the range of flat anomalous dispersion wavelength. Seven wavelength-paired photon pairs are configured as multi-wavelength energy-time entangled photon pair source and heralded single-photon source, respectively. A typical raw two-photon interference visibility of 95.5$\pm$6.5\% and a typical heralded second-order auto-correlation $g^{(2)}_{H}(0)$ of 0.045$\pm$0.001 are obtained. Our results pave the way for developing a fully integrated quantum photonic circuit.
\begin{figure}
    \centering
    \includegraphics[width=8.5 cm]{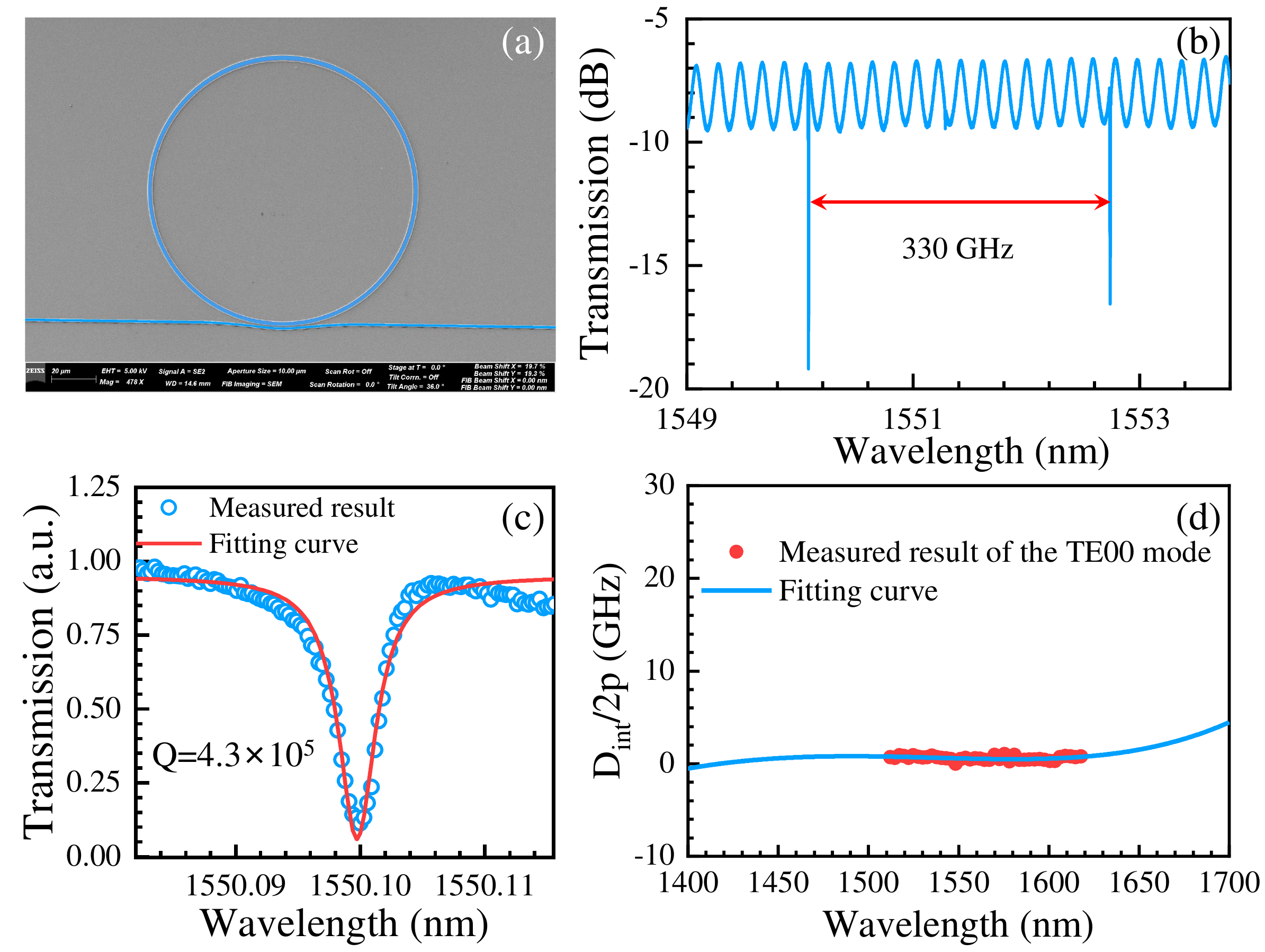}
    \caption{(a) Scanning electron microscopy image of the GaN MRR pulley with a 60-$\mu$m radius. (b) Measured transmission spectrum near 1550 nm with a free spectral range of 330 GHz. (c) Resonance dip around 1550.1 nm, indicating a loaded quality factor of $4.3\times10^5$. (d) Measured and fitted results of the integrated dispersion of the TE00 mode.}
    \label{fig:Fig1}
\end{figure}

\begin{figure}
    \centering
    \includegraphics[width=8.5 cm]{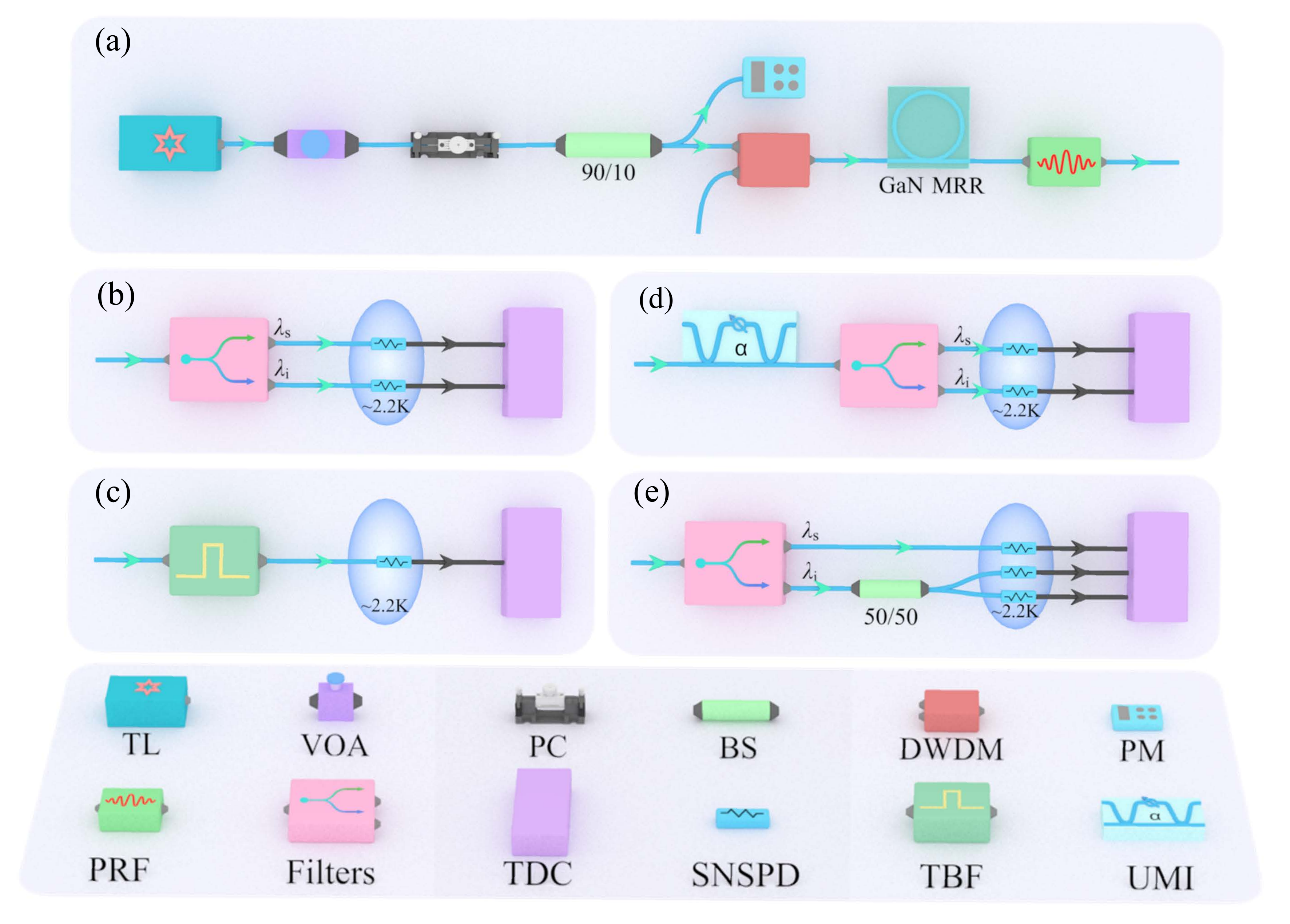}
    \caption{Schematic diagram of experimental setups. (a) Generation of correlated photon pairs. (b) Correlation properties. (c) Photon spectrum. (d) Energy-time entanglement with two-photon interference. (e) Heralded single-photon with HBT experimental setup. TL: tunable laser, VOA: variable optical attenuator, PC: polarization controller, BS: beam splitter, PM: power meter, DWDM: dense wavelength division multiplexer, PR: pump rejection; SNSPD: superconducting nanowire single-photon detector, TDC: time-to-digital converter, TBF: tunable bandpass filter, UMI: unbalanced Michelson interferometer. $\lambda_{s}$ and $\lambda_{i}$ are wavelengths of signal and idler photons, respectively. The SNSPDs are operated at a temperature of 2.2 K.}
    \label{fig:Fig2}
\end{figure}
\textit{Device fabrication and characterization.}—In our experiments, the MRR is fabricated on an undoped GaN film grown epitaxially via metal-organic chemical vapor deposition (MOCVD)\cite{Zheng2021GaN}. The radius is 60 $\mu$$m$ with an FSR of 330 GHz as shown in Fig.~{\ref{fig:Fig1}}(a) and (b).~Figure~{\ref{fig:Fig1}}(c) gives the transmission spectrum with a quality factor (Q-factor) of $4.3\times10^5$ at a resonant wavelength of 1550.1 nm. To obtain devices with anomalous and near-zero dispersion for the phase matching of SFWM, we simulate and design the GaN microring with 2.25-$\mu$$m$ waveguide width and 0.73-$\mu$$m$ etching depth. The experimental measured and fitted dispersions of the TE00 mode are shown in Fig.~{\ref{fig:Fig1}}(d), indicating an anomalous and near-zero dispersion of $-8.26\times10^{-27} s^2/m$ in a wide spectrum. See more details of device design and fabrication in Supplementary Materials Note1.

\begin{figure*}
    \centering
    \includegraphics[width=17.7 cm]{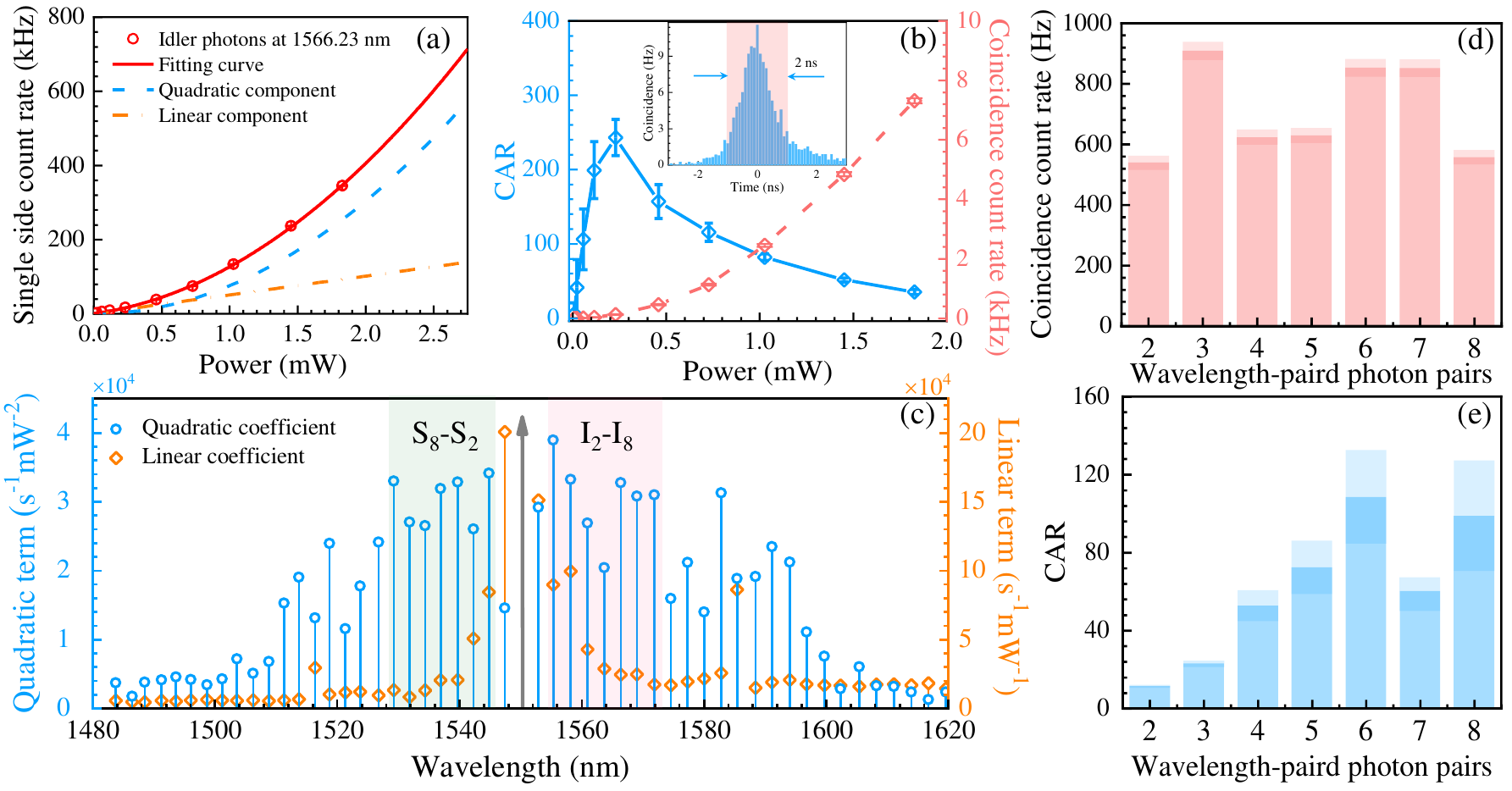}
    \caption{Experimental results of generated correlated photon pairs. (a) Single side count rate of the idler photon at 1566.23 nm versus pump power. (b) Coincidence count rate and the calculated CAR versus pump power. The inset is the measured coincidence histogram of the signal and idler photons at 1534.30 nm and 1566.23 nm. (c) Spectra of the correlated photons and noise photons in the on-resonance case from 1480 nm to 1620 nm. (d) Coincidence count rate of different combinations of the wavelength of correlated photon pairs. (e) CAR of different combinations of wavelength of correlated photon pairs.}
    \label{fig:Fig3}
\end{figure*}

\textit{Photon pair generation in the GaN MRR.}—The scheme for the generation and characterization of photon pair is illustrated in Fig.~{\ref{fig:Fig2}}. Figure~{\ref{fig:Fig2}}(a) shows the experimental setup for the generation of correlated photons in GaN MRR. In our work, we use a continuous-wave (CW) tunable laser (TL) at a wavelength of 1550.1 nm corresponding to the ITU channel 34 (C34). The power and polarization state of the pump light are adjusted by using a variable optical attenuator (VOA) and a polarization controller (PC), respectively. A 90:10 beam splitter (BS) is used for the power monitor with a power meter (PM). To suppress the sideband noise of the pump laser and the Raman photons generated in the fiber pigtails, a high-isolation ($\ge$120 dB) dense wavelength division multiplexer (DWDM) at C34 with a 20-cm long lensed fiber pigtail is connected to the input port of the chip. At the output of the chip, the residual pump laser is eliminated by a pump rejection (PR) filtering module with an isolation of $\ge$50 dB, which is coupled to the chip with another 20-cm long lensed fiber pigtail. An input-to-output coupling loss of 8.0 dB is achieved in our experiment. The generated photon pair, i.e., signal and idler photons are selected by another two DWDMs and detected by superconducting nanowire single-photon detectors (SNSPDs) with a detection efficiency of 75\% and a dark count rate of 80 Hz as shown in Fig.~{\ref{fig:Fig2}}(b). The detection signals from SNSPDs are sent to a time-to-digital converter (TDC) to record coincidence events. 

We measure the single side count rate and the coincidence count rate at different pump power levels to characterize the quantum correlation property of generated photons.~Figure~{\ref{fig:Fig3}}(a) shows the measured single side count rate of the idler photon at the wavelength of 1566.23 nm (red dot) as a function of pump power. The error bars of the photon count rate are obtained by Poissonian photon-counting statistics. The generation of correlated photon pairs is verified by fitting the experimental result with $N=a\times P+b\times P^{2}+c$, where $a$, $b$, and $c$ are the contributions of noise photon (yellow dash line), correlated photon (blue dash line), and dark count, respectively. By extracting the coefficient of the quadratic fitting curve, we obtain $a=5.1\times 10^{4}~s^{-1} mW^{-1}$ and $b=7.6\times10^{4}~s^{-1} mW^{-2}$, indicating the high-quality generation of correlated photon pair in our experiment. The coincidence count rate and coincidence-to-accidental ratio (CAR) are measured as shown in Fig.~{\ref{fig:Fig3}}(b) with the signal and idler photons at 1534.30 nm and 1566.23 nm, respectively. The average CAR reaches 243 with a detected coincidence count rate of 126 Hz with a coincidence width of 2 ns as illustrated in the inset of Fig.~{\ref{fig:Fig3}}(b). The efficiency or brightness ($B$) for single photon generation, i.e., photon pair generation rate ($PGR$) over on-chip pump power ($B=PGR/P_{p}^{2}$), is 2.09 $MHz\cdot mW^{-2}$ in average \cite{ma2020ultrabright}. Furthermore, we demonstrate the multi-wavelength property of the generated quantum light as shown in Fig.~{\ref{fig:Fig2}}(c).~The spectra of correlated and noise photons from 1480 nm to 1620 nm on each resonance are measured with a tunable bandpass filter (TBF, EXFO XTA-50) in Fig.~{\ref{fig:Fig3}}(c).~Our results show that multi-wavelength correlated photon pairs are generated exceeding a spectral range of 100 nm. Two peaks of noise photons appear at 1516.5 nm and 1585.3 nm due to the spontaneous Raman scattering of GaN\cite{harima2002properties,kuball2001raman}. See more details in Supplementary Materials Note2.~The quantum correlation properties of these photon pairs are characterized by measuring the coincidence events between different wavelength of signal/idler photon, i.e., $S_{i}I_{i}$, where $i$ is 2, 3, ...8 as the shaded area in Fig.~{\ref{fig:Fig3}}(c). Note that the result of the first wavelength-paired photon pair and the ones out of the range of our DWDMs are not given\cite{Pfeiffer2018, Wu2020, hance2021backscatter}. See more details in Supplementary Materials Note1. Figure~{\ref{fig:Fig3}}(d) shows the coincidence count rate of seven wavelength-paired resonances with a pump power of 1.1 mW. The CARs for particular pairs are illustrated in Fig.~{\ref{fig:Fig3}}(e), which gives a maximum average CAR of 108 with a coincidence count rate of 853 Hz.

\begin{figure}
    \centering
    \includegraphics[width=8.5 cm]{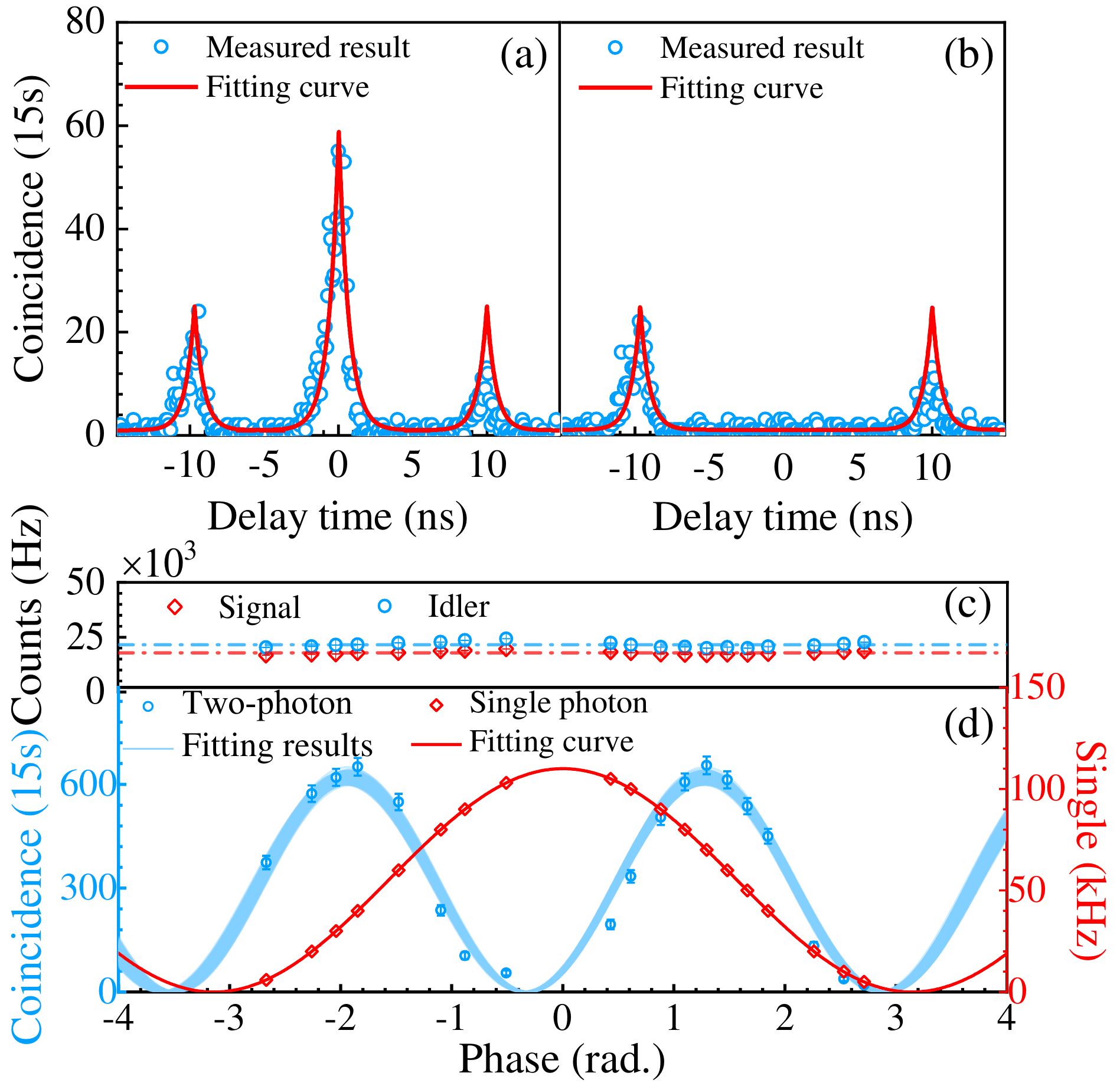}
    \caption{Experimental results of two-photon interference. (a) and (b) correspond to constructive and destructive two-photon interference within 15 s, respectively. (c) Photon count rate of the signal and idler photons. (d) Interference fringe of energy-time entangled photons with a visibility of 95.5±6.5\% (blue dots and lines). The single-photon interference is given by the right vertical axis with red dots and line.}
    \label{fig:Fig5}
\end{figure}

\begin{table*}
\caption{\label{tab:tableI} Results of visibilities of two-photon interference, $g^{(2)}(0)$, and $g^{(2)}_{H}(0)$ for the correlated photon pairs at different wavelengths.}
\begin{ruledtabular}
\begin{tabular}{ccccc}
$\lambda_{s}$ and $\lambda_{i}$ (nm) & Visibility(\%)&$g^{(2)}(0)$& $g^{(2)}_{H}(0)$ &Heralding count rate\\\hline
1544.80 $\&$ 1555.44 &82.3$\pm$4.1& 1.799$\pm$0.032 & 0.239$\pm$0.008 & 303 kHz\\
1542.16 $\&$ 1558.13 &82.9$\pm$1.3& 1.970$\pm$0.034 & 0.167$\pm$0.004 & 255 kHz\\
1539.53 $\&$ 1560.82 &96.6$\pm$2.0& 1.820$\pm$0.046 & 0.073$\pm$0.003 & 169 kHz\\
1536.91 $\&$ 1563.52 &99.3$\pm$4.9& 1.914$\pm$0.058 & 0.056$\pm$0.003 & 137 kHz\\
1534.30 $\&$ 1566.23 &95.5$\pm$6.5& 1.963$\pm$0.045 & 0.045$\pm$0.001 & 189 kHz\\
1531.70 $\&$ 1568.96 &94.3$\pm$5.6& 1.589$\pm$0.031 & 0.057$\pm$0.002 & 186 kHz\\
1529.11 $\&$ 1571.69 &99.2$\pm$5.7& 1.813$\pm$0.051 & 0.047$\pm$0.002 & 158 kHz\\
\end{tabular}
\end{ruledtabular}
\end{table*}

\textit{Energy-time entanglement.}—The correlated photon pairs generated in the MRR pumped by a CW laser have the intrinsic property of energy-time entanglement.~As shown in Fig.~{\ref{fig:Fig2}}(d), the quantum entanglement property in our experiments is verified by the two-photon interference \cite{franson1989bell, tittel1999long} in an unbalanced Michelson interferometer (UMI) with a 10-ns delay. The coincidence histograms after interference are shown in Figs.~{\ref{fig:Fig5}}(a) and (b), which represent the constructive and destructive two-photon interfering, respectively.~As shown in the Fig.~{\ref{fig:Fig5}}(c), count rates of signal and idler photons are constant at $\sim$17.8 kHz and $\sim$21.5 kHz, respectively, which indicates that there is no single-photon interference in our measurement. The measured interference curve is shown in Fig.~{\ref{fig:Fig5}}(d). Blue circles are experimental results, while the blue lines are the fitting curves with a 1000-time Monte Carlo method. The fitting visibility is calculated as 95.45±6.46\% without subtracting the accidental coincidence counts. An attenuated CW laser is injected into the UMI in the opposite direction for actively stabilization of the phase of the UMI. The measured single-photon interference of the attenuated laser is illustrated by red dots and line as shown in Fig.~{\ref{fig:Fig5}}(d). It can be seen that the period of single-photon interference is two times of that of the two-photon interference, verifying the energy-time property of the generated photon pairs. The measured energy-time entanglement properties of other wavelength-paired photon pairs are given in Table~{\ref{tab:tableI}}.  Due to the smaller CAR caused by extra noise photons, the visibilities of the second and the third wavelength-paired photon pairs are about 82\%, while the visibilities of the others are above 94\%.

\begin{figure}
    \centering
    \includegraphics[width=8.5 cm]{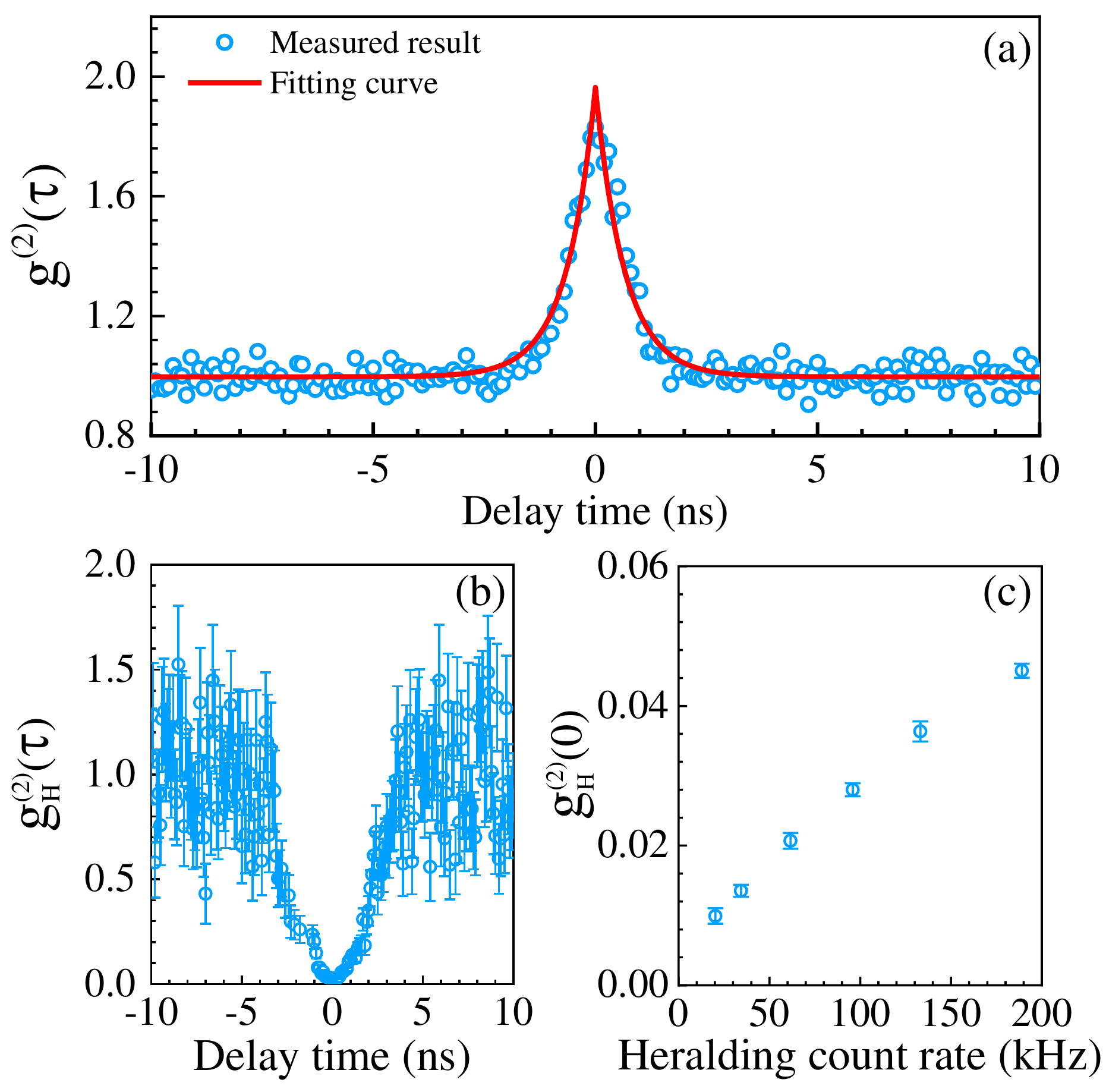}
    \caption{Characterization of single-photon purity. (a) Unheralded second-order auto-correlation coincidence histogram of photon at 1566.23 nm. (b) Heralded second-order auto-correlation $g^{(2)}_{H}(\tau)$ of heralded photon at 1566.23 nm and heralding photons at 1534.30 nm. (c) Heralded second-order auto-correlation $g^{(2)}_{H}(0)$ versus heralding count rate. }
    \label{fig:Fig4}
\end{figure}

\textit{Single-photon purity.}—In our demonstration, heralded single photons at different wavelengths can be obtained based on the generation of correlated photon pairs. The unheralded second-order auto-correlation function $g^{(2)}(\tau)$ is measured by the Hanbury Brown and Twiss (HBT) setup\cite{brown1954lxxiv} as shown in Fig.~{\ref{fig:Fig2}}(e). To characterize the single-mode property of generated photons on each resonance, the photons are sent into a 50:50 BS and the two-fold coincidence events are recorded. For an ideal single-mode thermal state, the $g^{(2)}(0)$ value should be 2, which can be obtained by calculating the ratio of the coincidence with zero delay to the one with nonzero delay. The measured result of auto-correlation $g^{(2)}(\tau)$ for photons at 1566.23 nm is shown in Fig.~{\ref{fig:Fig4}}(a). The measured data is fitted with a double exponential curve. The $g^{(2)}(0)$ of 1.963±0.045 is obtained, which corresponds to an effective mode number of 1.038 calculated by $g^{(2)}(0)=1+1/N$, where N is the total number of modes\cite{mandel1958fluctuations,caspani2017integrated}. The values of $g^{(2)}(0)$ for all the seven measured channels are given in Table~{\ref{tab:tableI}}. The heralded second-order auto-correlation function $g^{(2)}_{H}(\tau)$ for single-photon purity is measured with three-fold coincidence configuration. In our experiment, signal photons are detected by SNSPD while the idler photons are detected with a delay time of $\tau$ after passing through the 50:50 BS. Then, the three-fold coincidence events are recorded by TDC. With a pump power of 1.45 mW, the measured $g^{(2)}_{H}(\tau)$ for heralded single photon at 1566.23 nm, with its heralding at 1534.30 nm, is given in Fig.~{\ref{fig:Fig4}}(b). The obtained value of $g^{(2)}_{H}(0)$ is 0.045$\pm$0.001 with a heralding rate of 189 kHz. The values of heralded $g^{(2)}_{H}(0)$ with different heralding rates are shown in Fig.~{\ref{fig:Fig4}}(c). The values of $g^{(2)}_{H}(0)$ for heralded single photons on seven resonances are shown in Table~{\ref{tab:tableI}}.

\textit{Discussions and summary.}—In this work, we show that the GaNOI provides an important possibility for the quantum photonic integrated circuit. On one hand, the fabricated device exhibits a near-zero and flat anomalous dispersion over a large wavelength range of more than 100 nm. This indicates a great potential for generating multiple wavelength-paired correlated photon pairs, which is a pivotal advancement towards large-scale quantum networks.~In our current demonstration, we do not yet exhaust the maximum number of the ring resonances, which could provide us with eighteen wavelength-paired correlated photon pairs and could be further increased by reducing the FSR of the MRR. On the other hand, the GaNOI holds considerable promise for all-on-chip quantum photonic integrated circuit compared to existing platforms.~The on-chip integration of the pump laser could be realized on GaNOI with optical gain and nonlinear optical process.~For instance, InGaN/GaN laser diodes operating at a wavelength from 360 nm to 520 nm have been demonstrated\cite{watson2018ingan}, which could be used for the generation of correlated photon pairs via spontaneous parametric down-conversion (SPDC) within 720 nm to 1040 nm. Besides, with a proper portion of Indium, the bandgap energy of InGaN can be controlled from 0.65 to 3.4 eV, corresponding to operating wavelength 365 nm–1900 nm\cite{ohkawa2012740}.~At the same time, the optical filters for pump noise rejection and photon pair selection can also be realized on the GaNOI. Besides, the GaNOI also allows lattice-matched epitaxial deposition of Nb(Ti)N films for the on-chip integration of SNSPD\cite{sam2014high, redaelli2016design}. In our demonstration, although the GaNOI platform has shown groundbreaking for the generation of quantum light, the Raman noise is also observed in our experiment.~This is mainly due to the defects from the lattice mismatching between the GaN layer and the AlN buffer layer\cite{Zheng2021GaN, he2023ultra} and should be further eliminated by growing thicker GaN film on the buffer layer. See more details in Supplementary Materials Note2. 

In summary, we have demonstrated the generation of correlated photon pairs via SFWM in a GaN MRR for the first time.~By leveraging our advances in compound semiconductor nanofabrication,~the GaN MRR with a Q-factor of 0.43 million is obtained with an FSR of 330 GHz. Correlated photon pairs are generated in a wavelength range over 100 nm with their quantum properties being characterized by the coincidence measurement,~the two-photon interference, and the HBT measurement.~A typical two-photon interference visibility of 95.5$\pm$6.5\% and heralded second-order auto-correlation $g^{(2)}_{H}(0)$ of 0.045$\pm$0.001 are obtained,~respectively.~These results show that the GaNOI platform has remarkable potential for the development of all-on-chip QPIC.

\begin{acknowledgments}
This work was supported by Sichuan Science and Technology Program (Nos.~2021YFSY0063, 2021YFSY0062, 2021YFSY0064, 2021YFSY0065, 2021YFSY0066, 2022YFSY0061, 2022YFSY0062, 2022YFSY0063), the National Natural Science Foundation of China (Nos.~92365106, 62005039, 91836102, U19A2076),~Innovation Program for Quantum Science and Technology (Nos.~2021ZD0300701, 2021ZD0301702).

\end{acknowledgments}

\bibliography{reference}
\end{document}


\preprint{APS/123-QED}

\title{Supplementary Materials:\\Quantum Light Generation based on GaN Microring towards Fully On-chip Source}

\maketitle
\begin{center}
\textbf{Note1.}
\textbf{Device fabrication and characterization}
\end{center}

\begin{figure}[h]
  \centering
    \includegraphics[width=8.5 cm]{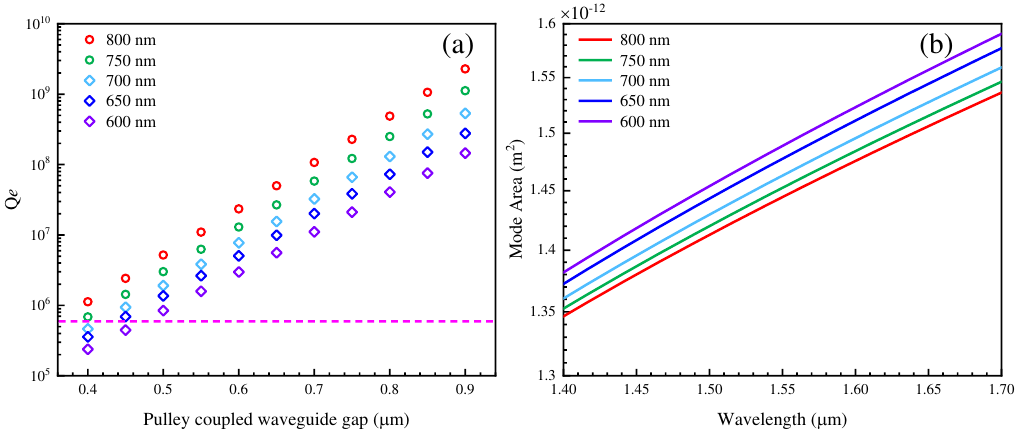}
    \caption{Simulated results for microring device. (a) Variation of $Q_e$ with pulley coupled waveguide gap, (b) variation of mode area with the wavelength at different etching depths.}
    \label{fig:SMFig.S0}  
\end{figure}

In our device design, the microring is expected to be over-coupled with a $Q_{e}/Q_{i}=3/5$ for the optimized generation and emission of correlated/entangled photon pairs, where $Q_e$ and $Q_i$ represent the external quality factor and the intrinsic quality factor, respectively\cite{fan2023multi}.~Figure~{\ref{fig:SMFig.S0}}(a) shows the simulated results of $Q_e$ with different pulley coupled waveguide gaps and different etching depths at a fixed $Q_i$ of $10^6$.~The optimal $Q_e$ is $6\times 10^5$ marked as the dashed line. The results indicate that the best choice is a waveguide gap of 400 nm with an etching depth of 750 nm or a waveguide gap of 450 nm with an etching depth of 650 nm.~However, in the process of our fabrication, limited by the proximity effect in electron beam lithography (EBL) and the lag effect of inductively coupled plasma (ICP) etching, achieving a device with a high Q-factor (\textgreater$10^5$) requires a gap wider than 500 nm. To increase the coupling efficiency, we choose the partial etching instead of the full etching, thus allowing a wider gap under the same coupling condition. As shown in Fig.~{\ref{fig:SMFig.S0}}(a), increasing the etching depth reduces the coupling coefficient and increases $Q_e$ under the same gap.~Meanwhile, increasing the etching depth leads to a reduced mode area, as shown in Fig.~{\ref{fig:SMFig.S0}}(b).~Therefore, we design the etching depth as 700 nm, which corresponds to a measured etching depth of 730 nm in our demonstration. Future optimization may involve adjusting the pulley waveguide to achieve phase matching, thereby increasing the gap between the ring and the pulley waveguide for over-coupling in photon pair generation.

\begin{figure}[h]
  \centering
    \includegraphics[width=8.5 cm]{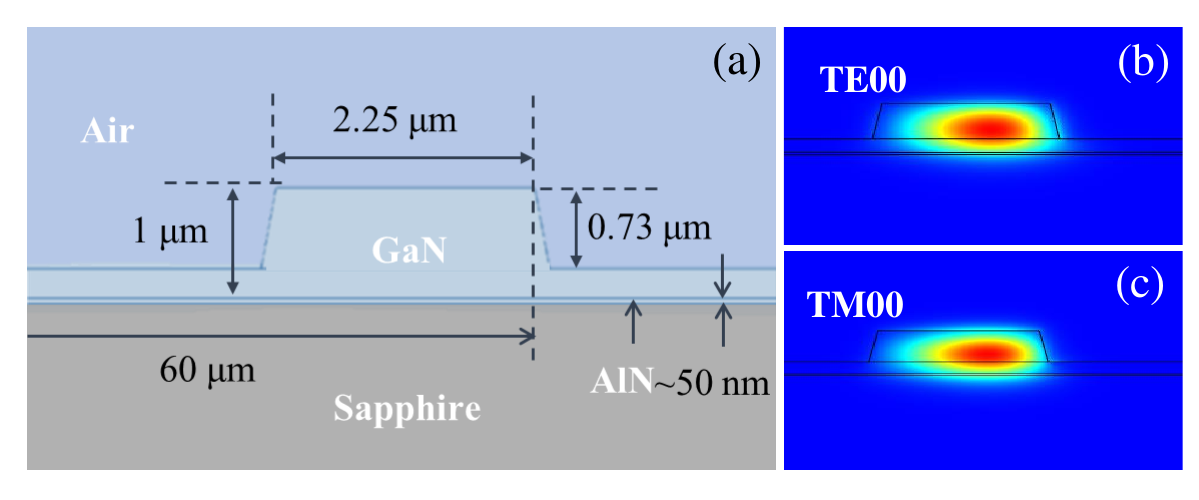}
    \caption{Device design. (a) GaN microring with a waveguide width of 2.25 $\mu$$m$ and an etching depth of
0.73 $\mu$$m$. (b) and (c) Simulated mode profiles of TE00 and TM00.}
    \label{fig:SMFig.S1}  
\end{figure}

The microring is manufactured on an undoped GaN film grown epitaxially via metal-organic chemical vapor deposition (MOCVD)\cite{zheng2022integrated}. The GaN film, with a thickness of 1 $\mu$m, is grown on a \textit{c}-plane sapphire substrate, atop which lies an approximately 50-nm-thick aluminum nitride (AlN) buffer layer. In this process, AlN film is formed by magnetron sputtering to improve the crystalline quality. Electron beam lithography (EBL) is utilized for patterning the microring and bus waveguide. Subsequently, the device undergoes a dry etching process using an inductively coupled plasma (ICP) process with a $Cl_2/Ar/BCl_3$ gas mixture. In our work, a GaN microring with a radius of 60 $\mu$$m$, a waveguide width of 2.25 $\mu$$m$, and an etching depth of 0.73 $\mu$$m$ is obtained, and the cross-section is shown in Fig.~{\ref{fig:SNFig.S1}}(a). The bus waveguide utilizes a pulley structure to enhance coupling efficiency, maintaining a 500-nm gap from the microring. The simulated mode profiles of TE00 and TM00 are shown in Fig.~{\ref{fig:SMFig.S1}}(b) and (c), demonstrating strong light confinement. 

\begin{figure}[h!]
  \centering
    \includegraphics[width=8 cm]{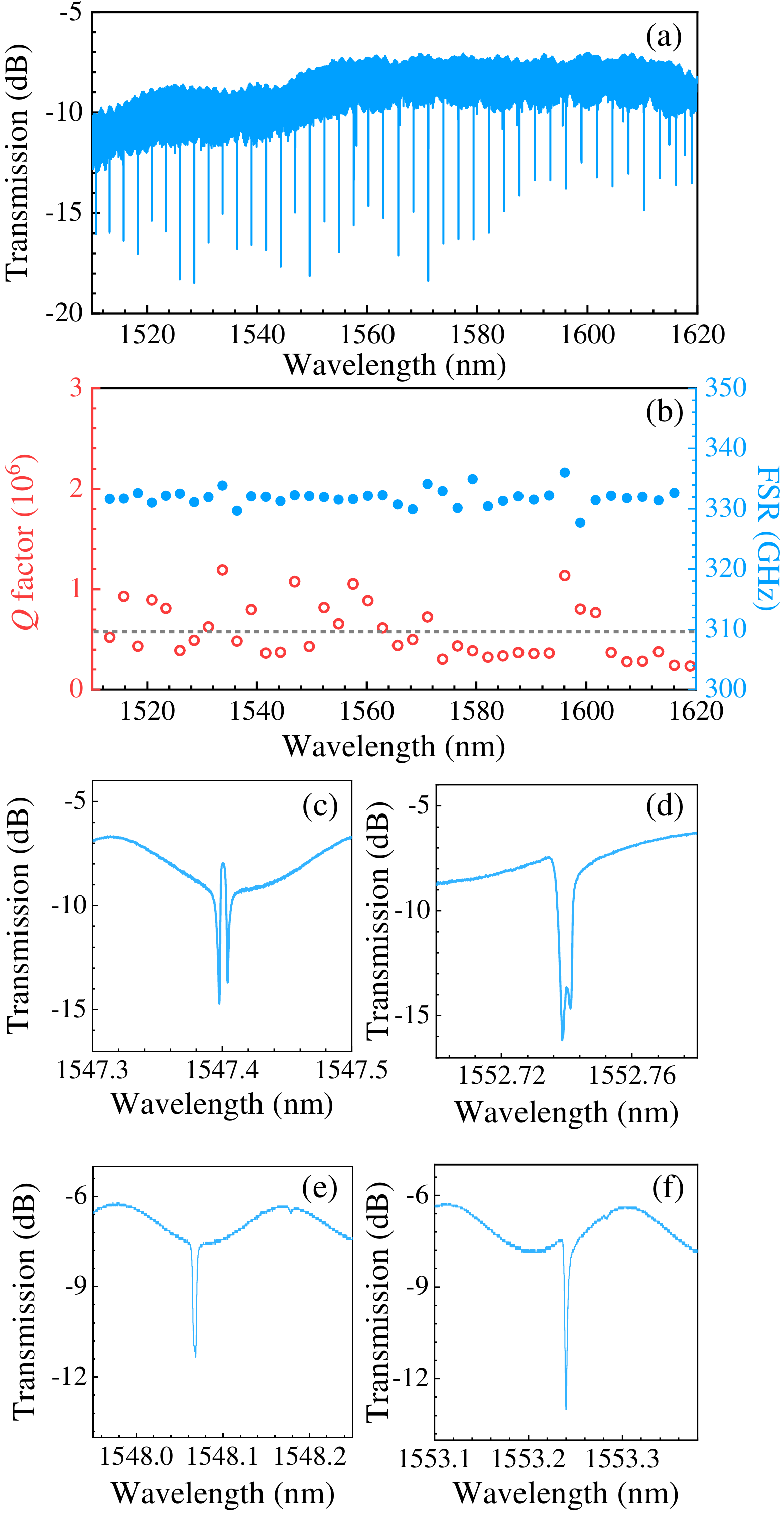}
    \caption{Device characterization. (a) Transmission spectrum from 1510 to 1620 nm with a free spectral range of 330 GHz. (b) Loaded Q-factors distribution with a maximum Q-factor of $1.2 \times 10^{6}$ and mean Q-factor of $5.8 \times 10^5$ - the dashed line - with an FSR of $\sim$330 GHz. (c) and (d) show the TE mode resonance dip splits at 1547.41 nm and 1552.74 nm, respectively. (e) and (f) show the TM mode resonance dip splits at 1547.41 nm and 1552.74 nm, respectively.}
    \label{fig:SMFig.S2}  
\end{figure}

The transmission property of the GaN MRR can be characterized by a tunable continuous-wave (CW) tunable laser, a photodetector, and an oscilloscope.~In Fig.~{\ref{fig:SMFig.S2}}(a), the transmission spectrum of the TE00 mode is measured from 1510 nm to 1620 nm. The loaded quality factors (Q-factors) and free spectral range (FSR) are analyzed, as depicted in Fig.~{\ref{fig:SMFig.S2}}(b), revealing a maximum Q-factor of $1.2 \times 10^{6}$ and a mean Q-factor of $5.8 \times 10^5$ - the dashed line - with FSRs of $\sim$330 GHz. It is noted that fundamental resonance doublet by backscatter effects are observed at 1547.41 nm and 1552.74 nm due to the surface roughness from the sidewalls, as shown in Figs.~{\ref{fig:SMFig.S2}}(c) and (d), respectively. As shown in Figs.~{\ref{fig:SMFig.S1}}(b) and (c), it can be seen that the light field of the TE00 mode overlaps more with the sidewalls compared to that of the TM00 mode, resulting in a higher backscatter coupling efficiency. The resonance doublet would be observed if the coupling rate to the counter-propagating mode is similar or larger than the total resonance decay rate\cite{Pfeiffer2018,Wu2020}. This is evident in Fig.~{\ref{fig:SMFig.S2}}(e) and (f), where the absence of a resonance doublet indicates a small coupling efficiency caused by scattering at the top of the waveguide for the TM00 mode. This feature diminishes the performance of the photon pair at these wavelengths in the main text\cite{hance2021backscatter}. In future work, we can optimize the inductively coupled plasma (ICP) etching conditions to achieve smooth sidewalls and mitigate mode splitting.
\begin{figure}
  \centering
    \includegraphics[width=8 cm]{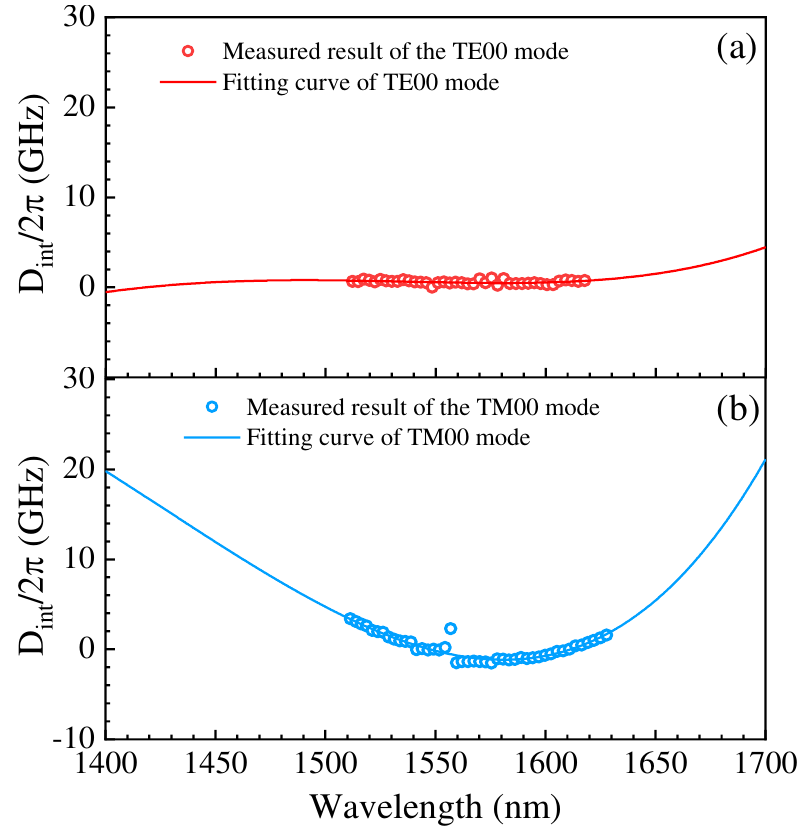}
    \caption{(a) Measured and fitted integrated dispersions of the TE00 mode, (b) measured and fitted integrated dispersions of the TM00 mode.}
    \label{fig:SMFig.S3}  
\end{figure}

The microring dispersion can be expressed as
\begin{equation}\label{...}
    \omega_{\mu} = \omega_0 + D_1\mu + \frac{1}{2!}D_2\mu^{2} + \frac{1}{3!}D_3\mu^{3} + \frac{1}{4!}D_4\mu^{4} + ...
\end{equation}
\begin{equation}\label{...}
    \omega_{\mu} = \omega_0 + D_1\mu + D_{int}
\end{equation}
where $\mu$ refers to the sequence number of resonant modes relative to the pump mode ($\mu$=0) and $D_{int}$ is the integrated dispersion of the microring. The measured and fitted results in Fig.~{\ref{fig:SMFig.S3}} show that TE00 mode exhibits a flatter and smaller anomalous dispersion compared to the TM00 mode, with a dispersion coefficient of $8.26\times10^{-27} s^2/m$ for the TE00 mode.
\begin{figure}[h]
  \centering
    \includegraphics[width=8 cm]{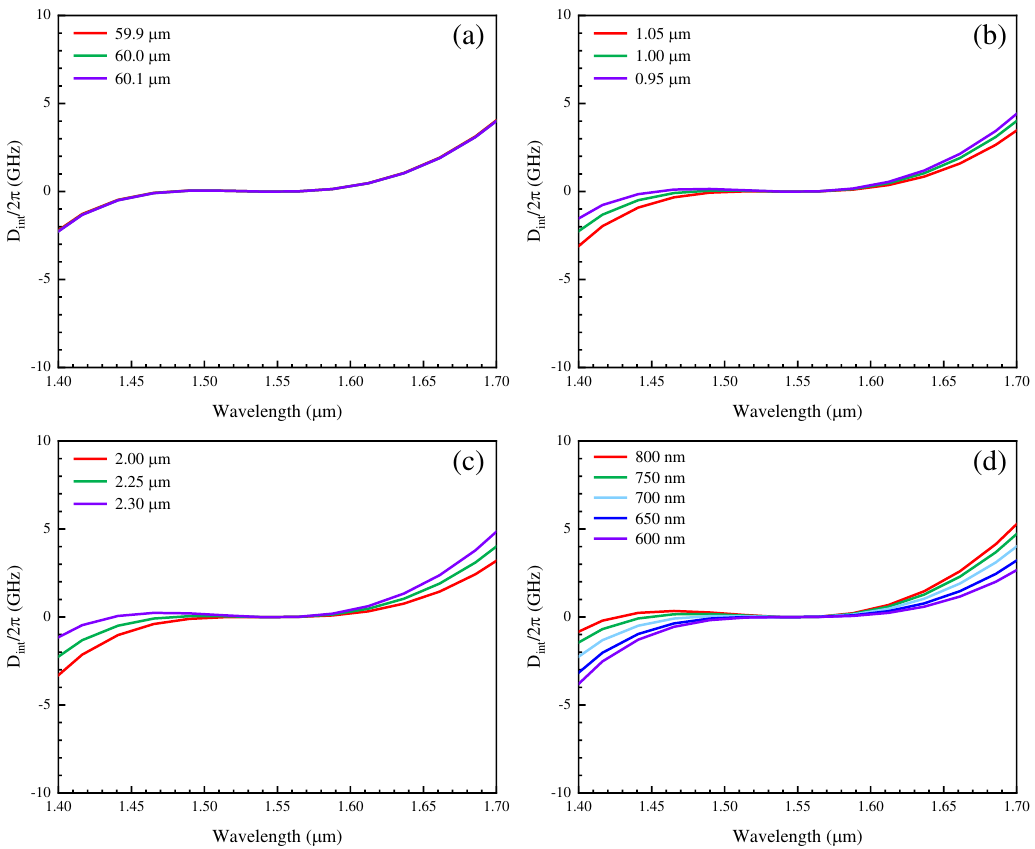}
    \caption{Simulated results of dispersion with different parameters. (a) Radius of microring, (b) thickness of waveguide, (c) width of waveguide, and (d) etching depth.}
    \label{fig:SMFig.S00}  
\end{figure}

Regarding the robustness of our design, the etching rate of ICP is approximately 270 nm/min with a slight fluctuation. In this case, the resolution of the etching depth can be controlled within 50 nm. To compensate for fabrication error, the actual device preparation gap starts from 400 nm and gradually changes with a step of 50 nm, allowing us to obtain devices with different coupling conditions. On the other hand, the robustness of our device can be also characterized by the dispersion. Figures~{\ref{fig:SMFig.S00}}(a)-(d) show the influence of fabrication errors in radius, thickness, width, and etching depth on the integrated dispersion of the TE00 mode. The results show that the dispersion remains flat and near-zero from 1500 nm to 1600 nm. It can be seen that the fabricated device is robust despite the variations between the fabrication and the design.

\begin{figure}[H]
  \centering
    \includegraphics[width=8 cm]{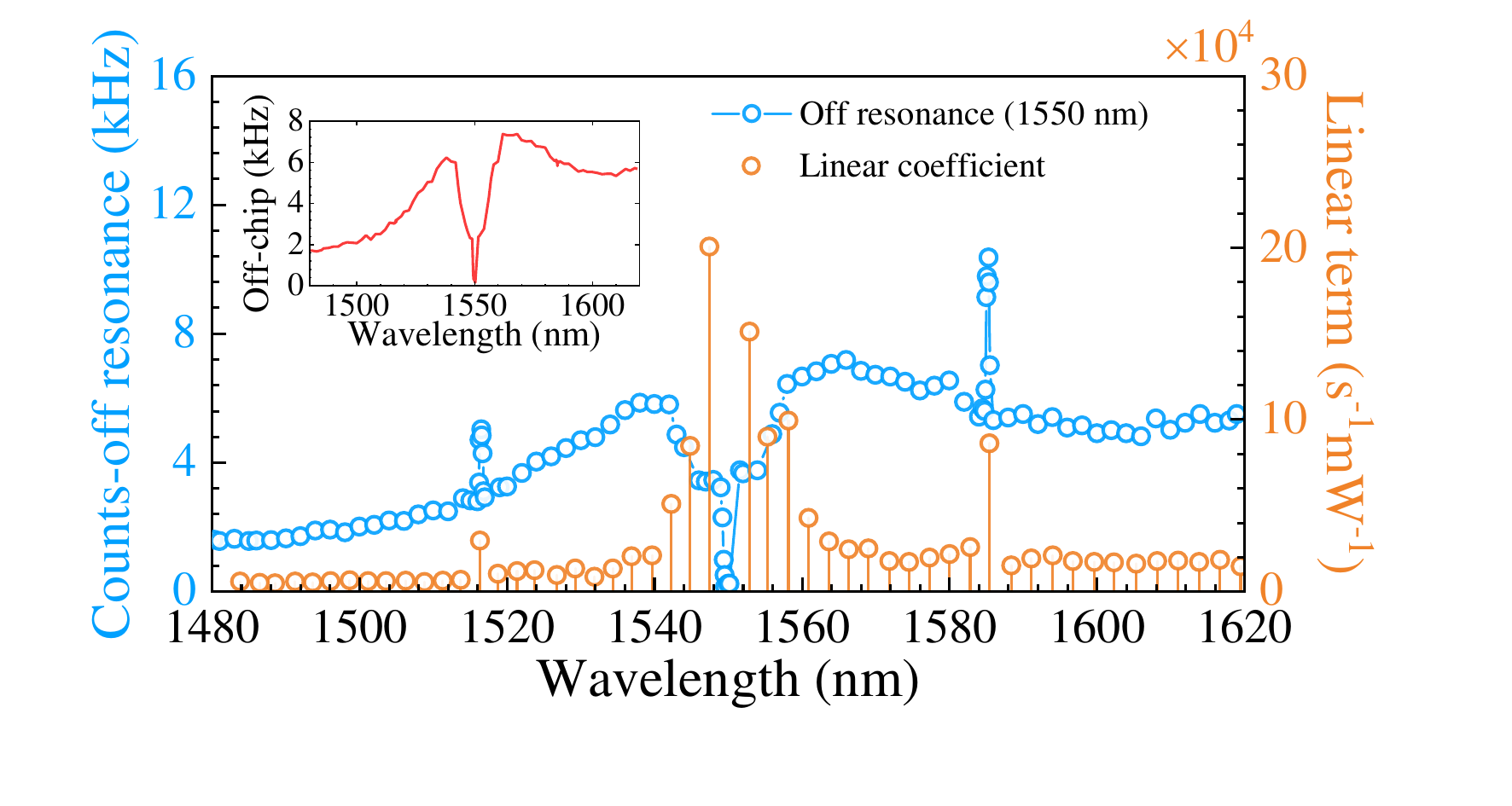}
    \caption{Spectral response of the GaN MRR for the off-resonance case.~The insert shows the spontaneous Raman noise generated by the experimental setup without the GaN MRR.}
    \label{fig:SMFig.S4}  
\end{figure}

\begin{center}
\textbf{Note2.}
\textbf{Spectra of noise and correlated photons}
\end{center}
To evaluate the generation of noise and correlated photons, the spectra are recorded by frequency scanning with a tunable bandpass filter (TBF, XTA-50, EXFO).~Firstly, we measure the Raman scattering noise photons of the optical fiber system, i.e. without the GaN MRR chip, as shown in the insert of Fig.~{\ref{fig:SMFig.S4}}, which is operated by directly aligning two lens fibers.~The dip near 1550.1 nm demonstrates the significant effectiveness of the high-rejection filters in pump rejection. Then, the noise photons with GaN chip in the off-resonance case are characterized in Fig.~{\ref{fig:SMFig.S4}} with blue circles, which are composed of photons in optical fiber optic systems and the GaN bus waveguide.~The measured spectral envelope exhibits two peaks at 1516.5 nm and 1585.3 nm, respectively, which corresponds to the $E_2$ (low) phonon mode frequency peak with Raman shift of 144 $cm^{-1}$ in the first-order Raman scattering spectrum of GaN\cite{harima2002properties,kuball2001raman}. Compared to the spectrum of noise photons in the on-resonance case, i.e., the linear coefficient, the Raman peaks in GaN coincide with the signal and idler wavelength channels, leading to a sudden surge in linear noise and a reduction in the signal-to-noise ratio of the light source. 

\begin{figure}[h]
  \centering
    \includegraphics[width=8 cm]{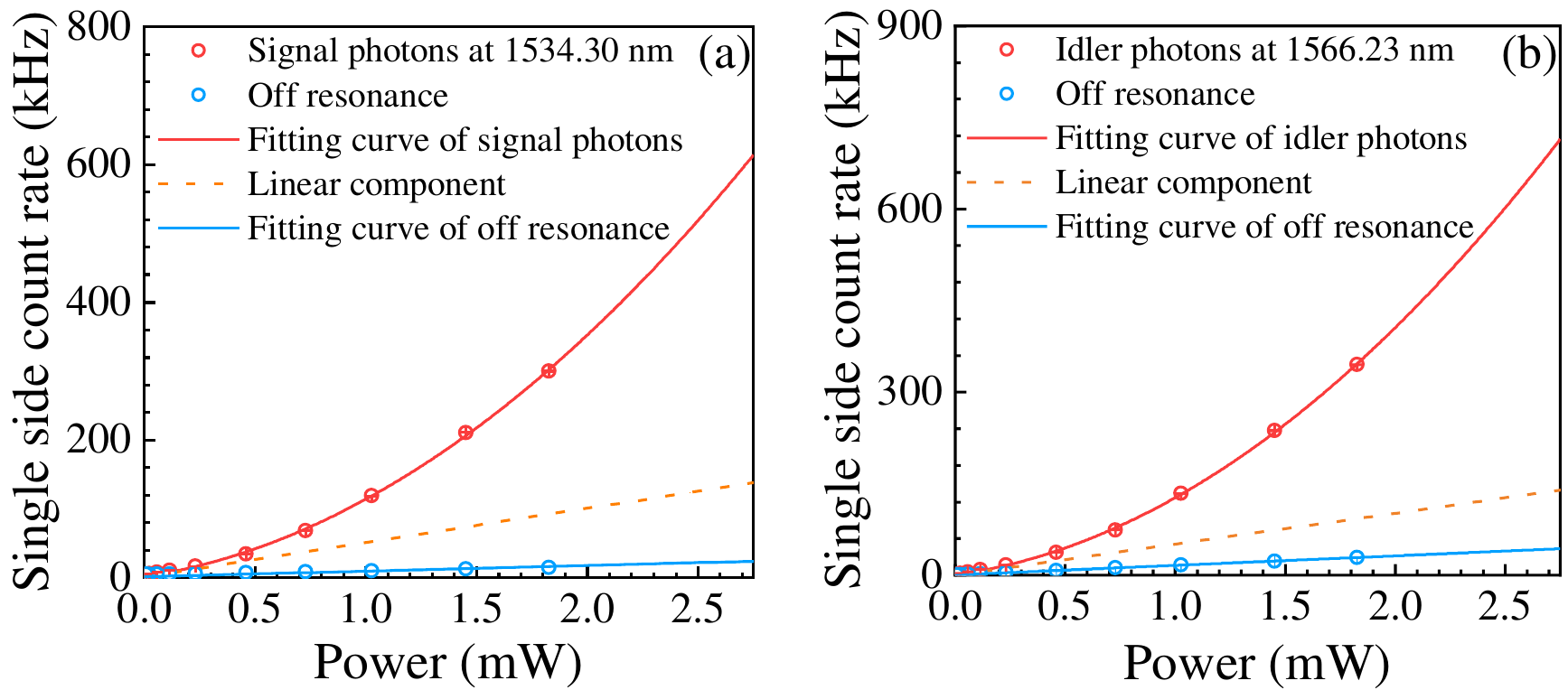}
    \caption{Noise photons from microring.~(a) and (b) Single side count rate at 1534.30 and 1566.23 nm for on-resonance and off-resonance cases, respectively.}
    \label{fig:SMFig.S5}  
\end{figure}

As shown in Fig.~{\ref{fig:SMFig.S5}}, the single side count rates of photons at 1534.30 and 1566.23 nm for on-resonance and off-resonance cases under different pump power levels are measured, respectively. The results of the on-resonance case are fitted with a quadratic polynomial formed by $N = a \times P + b \times P^2 + c$ to verify the generation of correlated photon pairs and noise. The linear component represents the presence of noise photons from the optical fiber, bus waveguide, and microring. The noise photon counts for the off-resonance case are proportional to the injected pump power. We observe that the level of noise photons in the off-resonance case is lower compared to the on-resonance case, accounting for 16\% and 31\% for idler and signal, respectively. This indicates that most of the noise photons are generated from the microring and are amplified by the enhancement effect of the cavity. The generation of the noise photons is due to the lattice mismatch between the GaN layer and the AlN buffer layer and could be further eliminated by growing thicker GaN film on the buffer layer.

\bibliography{supplements}